\documentclass[latterpaper,12pt]{article}
\pdfoutput=1

\usepackage{amsmath}
\usepackage{amsfonts}
\usepackage{amssymb}
\usepackage{graphicx}
\usepackage{color}
\usepackage{slashed}
\usepackage{dsfont}
\usepackage{bbold}
\usepackage{color}

\usepackage{graphicx}
\usepackage[body={6.5in, 9in}, right=1in, top=1in]{geometry}
\usepackage[numbers,compress]{natbib}

\usepackage{amscd, latexsym}
\usepackage[mathcal]{euscript}
\usepackage{amssymb, epsfig}
\usepackage{mathrsfs}
\usepackage{a4wide}
\usepackage{enumerate}
\usepackage{mathtools}
\usepackage{subfigure}
\usepackage{verbatim}

\usepackage{pstricks}                                      %PS tricks packages
\usepackage{pst-node}
\usepackage{pst-tree}
\usepackage{pst-plot}
\usepackage{float}
\numberwithin{equation}{section}

\newcommand{\pderiv}[2]{\frac{\partial #1}{\partial #2}}

\newcommand{\pderivp}[2]{\frac{\partial }{\partial #2} \left(#1\right)}

\newcommand{\paren}[1]{\left( #1 \right)}

\newcommand{\be}{\begin{equation}}
\newcommand{\ee}{\end{equation}}

\newcommand{\Comment}[1]{{}}
\definecolor{MyDarkBlue}{rgb}{0.15,0.15,0.45}
\usepackage[linktocpage=true]{hyperref}
\hypersetup{
colorlinks=true,
citecolor=MyDarkBlue,
linkcolor=MyDarkBlue,
urlcolor=MyDarkBlue,
}

\begin{document}
\def\thefootnote{\fnsymbol{footnote}}

\begin{center}
\Large{\textbf{Leading slow roll corrections to the volume of the universe and the entropy bound}} \\[0.5cm]

\large{Matthew Lewandowski$^{\rm a}$, Ashley Perko$^{\rm a}$}
\\[0.5cm]

\small{
\textit{$^{\rm a}$ Stanford Institute for Theoretical Physics, Stanford University, Stanford, CA 94306}}

\end{center}

\vspace{.8cm}

\hrule \vspace{0.3cm}
\noindent \small{\textbf{Abstract}\\ We make an extension to recent calculations of the probability density $\rho(V)$ for the volume of the universe after inflation.  Previous results have been accurate to leading order in the slow roll parameters $\epsilon \equiv \dot{H}/H^2$ and $\eta \equiv \ddot{\phi}/(\dot{\phi} H)$, and $1/N_c$, where $H$ is the Hubble parameter and $N_c$ is the classical number of $e$-foldings.  Here, we present a modification which captures effects of order $\epsilon N_c$, which amounts to letting the parameters of inflation $H$ and $\dot{\phi}$ depend on the value of the inflaton $\phi$.  The phase of slow roll eternal inflation can be defined as when the probability to have an infinite volume is greater than zero. Using this definition, we study the Laplace transform of $\rho ( V )$ numerically to determine the condition that triggers the transition to eternal inflation.  We also study the average volume $\langle V \rangle$ analytically and show that it satisfies the universal volume bound.  This bound states that, in any realization of inflation which ends with a finite volume, an initial volume must grow by less than a factor of $e^{S_{\rm{dS}}/2}$, where $S_{\rm dS}$ is the de Sitter (dS) entropy. }
\vspace{0.3cm}
\noindent
\hrule
\def\thefootnote{\arabic{footnote}}
\setcounter{footnote}{0}

%%%%%%%%%%%%
%
%    introduction 
%
%
%%%%%%%%%%%%%%%%%

\section{Introduction} 
The classical theory of a slowly rolling scalar field $\phi$ minimally coupled to gravity describes an inflating space-time with scale factor $a(t) = e^{H (t - t_0 )}$.  Thus, a portion of space which has volume $V_0$ at time $t_0$ will exponentially expand to have a volume equal to $e^{ 3 H ( t - t_0) } V_0$ at a later time $t$.  Quantum fluctuations, however, change the picture.  If we describe inflation as ending at some value of the field $\phi_r$, then in order to calculate the volume of the universe after inflation, we want to calculate the volume of the reheating surface, i.e. all points $\mathbf{x}$ such that  $\phi( \mathbf{x} , t_r) = \phi_r$.  One could imagine doing this perturbatively by calculating the quantum fluctuation $\zeta = \delta a / a$ which describes the deformation of surfaces of constant $\phi$.  This approach fails if $\zeta$ is of order one, but this is precisely the regime of interest in this paper.  As we will discuss below, recent work \cite{Dubovsky2009, Dubovsky2011} has taken a different approach and resummed the quantum corrections and avoided this perturbative expansion.  The calculation is accurate to lowest order in the slow roll parameters $\epsilon \equiv \dot{H}/H^2$ and $\eta \equiv \ddot{\phi}/(\dot{\phi} H)$, and $1/N_c$, where $H$ is the Hubble parameter and $N_c$ is the classical number of $e$-foldings.  In this paper, we will find the corrections of order $\epsilon N_c$ to the analysis done in \cite{Dubovsky2009, Dubovsky2011}.  Our results are not an approximation for small $\epsilon N_c$, i.e. we are resumming these corrections.  We will see in Section \ref{review} that $\epsilon N_c$ cannot be larger than order one because of the universal volume bound, which will also be described below.  For the rest of the paper, we will use $\epsilon$ to represent both slow roll parameters, or, if you will, the larger of the two.  

In addition, \cite{Creminelli2008, Dubovsky2009} have given a strict criterion for when this system enters the eternal inflation regime, which was possible because the method accommodated large fluctuations $\zeta$.  Eternal inflation is a phase of inflation where the probability to have an infinite volume is larger than zero.  What parameter should describe the transition between slow roll inflation that eventually ends and eternal inflation?  Let's consider the relative importance of the classical and quantum motions of the inflaton $\phi$.  In a time $H^{-1}$, the field classically rolls a distance $\Delta \phi_{\rm{classical}} = \dot{\phi} H^{-1}$.  The field, however, has quantum jumps which are given by $\Delta \phi_{\rm{quantum}} \equiv \sqrt{ \langle \delta \phi^2 \rangle_H } \sim H$, where all relevant mass scales are evaluated at $H$.  Thus, we expect that the quantum motion should dominate if $\Delta \phi_{ \rm{classical}} / \Delta \phi_{ \rm{quantum}} \lesssim 1$.  While the classical motion always takes the field down the potential toward the reheating surface, quantum jumps can take it either up or down the potential.  If the quantum motion dominates, there may always be patches of space with field values which do not reach the reheating surface.  This is the eternal regime.  More specifically, \cite{Creminelli2008} found that the phase transition to eternal inflation occurs sharply when $\Omega \equiv 2 \pi^2 \dot{\phi}^2 / (3 H^4)$ becomes less than one. 

The analysis of  \cite{Dubovsky2009}, and a generalization to single field inflation in arbitrary dimensions, to multifield inflation, and when considering higher order derivative corrections (equivalent to $H/M_{Pl}$ corrections) \cite{Dubovsky2011}, has supported the validity of the universal volume bound.  The statement of the bound is that the maximum possible finite volume produced in a region of space which globally exits inflation is $e^{S_{\rm{dS}}/2}$ times the original volume.  Here, $S_{\rm dS} = 8 \pi^2 M_{Pl}^2 / H^2$ is the entropy of de Sitter space.  This factor is larger than the one found in \cite{Arkani-hamed}, $e^{S_{\rm dS}/4}$, which was calculated in the classical limit.  The existence of this bound has been compared to a similar bound in the context of black hole complementarity \cite{Verlinde1995, Lowe1995}.  In this paper, we show that effects of order $\epsilon N_c$ do not violate the full quantum bound $e^{S_{\rm{dS}}/2}$ \cite{Dubovsky2009, Dubovsky2011}, but always decrease the average volume so that the bound is more easily satisfied. 

This paper is organized as follows.  In Section \ref{review}, we will review the results of \cite{Dubovsky2009} to define our system.  In Section \ref{revisedbacteriamodel}, we will discuss our extension of that procedure which includes corrections of order $\epsilon N_c$ and in Section~\ref{orderofapprox} point out its limitations.  Then in Section~\ref{transition} we will present our results for the transition condition and in Section \ref{averagevolsec} the average volume, and show that it is consistent with the universal bound.  We work in the context of standard $3+1$ dimensional slow roll inflation in Einstein-Hilbert gravity, and use the conventions $8 \pi G_N = M_{Pl}^{-2}$, so that the Friedmann equation is $H^2 = ( \frac{1}{2} \dot{\phi}^2 + U(\phi) )/(3 M_{Pl}^2)$ for a potential $U(\phi)$.

%%%%%%%%%%%%%
%
%        setup
%
%
%%%%%%%%%%%%%

\section{Setup}

%********************%
%
%          review of previous results
%
%
%%%%%%%%%%%%%%

\subsection{Review of Previous Results} \label{review}

The authors of \cite{Dubovsky2009} calculated the probability distribution of the total volume of the universe, smoothed with a UV cutoff $\Lambda \ll H$, at the end of inflation to lowest order in the slow roll parameters and $1/N_c$.  They did this by considering a specific UV model, a bacteria model, which has the same IR properties as inflation. Under such conditions, the quantum field theory of the inflaton is equivalent to a classical stochastic system.  This means that all expectation values can be written as an integral over a classical probability distribution.  To this end, the authors of \cite{Dubovsky2009} were able to calculate $\rho ( V , \phi)$, the probability distribution for the total volume of the universe at the end of inflation as a function of starting position $\phi$, by expressing it as a Laplace transform
\be \label{laplacetrans}
\rho(V, \phi) = \frac{1}{2 \pi i} \int_{0^+ - i \infty}^{0^+ +i \infty} dz \, f( \phi ; z) e^{z V} \ .
\ee
The volume $V$ above is a dimensionless volume, the physical volume divided by the UV smoothed initial volume $H^{-3} (H / \Lambda)^3$.  The function $f( \phi ; z)$ satisfies
\be \label{apples}
\frac{1}{2} \frac{\partial^2 f}{\partial \phi^2} - \frac{2 \pi \sqrt{ 6 \Omega }}{H} \pderiv{f}{\phi} + \frac{12 \pi^2}{H^2} f \log f = 0 \ ,
\ee
with boundary conditions $f(\phi_r ; z) = e^{-z}$ and $\partial f / \partial \phi ( \phi_b) = 0$.  Here, $\phi_r$ is the field value of the reheating surface on which inflation ends, and $\phi_b$ is a reflective barrier.  This defines the range $( \phi_r , \phi_b)$ within which the inflaton rolls.  More details about the origin of this equation are given in Section \ref{revisedbacteriamodel}.

The function $f$ can be used to study the transition to eternal inflation.  The moments of the probability distribution can be calculated directly from (\ref{laplacetrans}): 
\be \label{moments}
\langle V^n(\phi) \rangle = \int_0^\infty V^n \rho( V , \phi) dV = (-1)^n \frac{ \partial^n f}{ \partial z^n}(\phi ; z \rightarrow 0) \ ,
\ee 
from which it follows that $f(\phi ; z \rightarrow 0 ) = \int \rho( V , \phi ) \, dV \equiv P_{\rm{finite}}$ is the probability to end inflation with a finite volume.  The authors of \cite{Dubovsky2009, Dubovsky2011} defined eternal inflation as the phase where $P_{\rm{finite}} < 1$, and they found that this occurred for $\Omega < 1$.  By inspection $f = 1$ is a solution to (\ref{apples}) regardless of the value of $\Omega$, so it seems like there is always a solution where $P_{\rm{finite}} = 1$.  However, the limiting procedure $z \rightarrow 0$ is important.  When $\Omega > 1$ there is a continuous family of solutions $f( \phi ; z )$ such that $f( \phi; z \rightarrow 0) = 1$, and therefore the system is not in the eternal regime.  However, when $\Omega < 1$, $f( \phi ;  z \rightarrow 0) \neq 1$ and so $P_{\rm{finite}} <1$ and the system is in the eternal regime.

Using equations (\ref{apples}) and (\ref{moments}) and letting $\dot{f} = \partial f / \partial z \equiv - \langle V \rangle  $ and $f' = \partial f / \partial \phi$, we find the differential equation for the average volume
\be \label{harrypotter}
\frac{1}{2} \dot{f}'' -  \frac{2 \pi \sqrt{ 6 \Omega }}{H} \dot{f}' +  \frac{12 \pi^2}{H^2} \dot{f} = 0 \ ,
\ee
where $\dot{f}$ satisfies the boundary conditions $\dot{f}(\phi_r ; 0 ) = -1$ and $\dot{f}'(\phi_b)=0$.  This equation is only valid before the transition to eternal inflation, when $f(\phi; z \rightarrow 0) = 1$ is a solution to (\ref{apples}).  The result found in \cite{Dubovsky2009} is
\be \label{oldvolume}
\langle V \rangle = e^{ 3 \frac{2}{1 + \sqrt{ 1 - 1 / \Omega} } N_c } \ .
\ee
The classical regime has $\Omega \gg 1$, so $\log \langle V \rangle \rightarrow 3 N_c$, which is indeed the classical value.  As the system becomes more quantum ($\Omega \rightarrow 1$) there is a huge increase in the volume.  Just at the transition $\Omega = 1$, the maximum volume is obtained: $\log \langle V \rangle = 6 N_c$.  It was shown in \cite{Dubovsky2009} that the maximum finite volume produced, even in the eternal regime, is $V_{\rm max} = \langle V \rangle_{\Omega = 1}$, up to non-perturbatively small corrections.  We can write $N_c$ in terms of the dS entropy as follows.  Keeping in mind the slow roll equation $\dot{\phi} = - 2 M_{Pl}^2 H'$, the definition $dN_c = H dt = (H / \dot{\phi}) d \phi$, and keeping $\Omega$ constant, the change in the entropy along a trajectory is $\Delta S_{\rm dS}  = \Delta \left( 8 \pi^2 M_{Pl}^2 / H^2 \right) = 8 \pi^2 \int (\dot{\phi}^2/H^4 ) dN_c = 12 \Omega N_c$.  Thus $V_{\rm max} = e^{S_{\rm dS } (\phi_r) /2}$ gives the universal bound on the volume.  

This puts a limit on how large $\epsilon N_c$ can be.  From the discussion above $ \Omega  N_c \leq S_{\rm ds}(\phi_r) /12$, and using $ \epsilon = - \dot \phi^2 /( 2 H^2 M_{Pl}^2) = - 6 \, \Omega / S_{\rm dS} $ 
\be
| \epsilon ( \phi_r) N_c | = 6 \, \Omega \frac{N_c}{S_{\rm dS} (\phi_r)} \leq \frac{1}{2} \ .
\ee
This is only valid for the constant $\Omega$ case, but shows that $\epsilon  N_c$ can generally be of order unity.

We should make one more comment about the volume that appears in equation (\ref{laplacetrans}) and throughout the rest of the paper.  It is the volume produced after starting with \emph{one} patch of space of volume $V_{\Lambda} \equiv H^{-3} (H/\Lambda)^3$.  With more than one starting patch, the distribution $\rho(V)$ will be different because there will be many possible ways of achieving the volume $V$ from the initial configuration.  However, for the purposes of this paper, the difference is inconsequential.  We focus on the average volume $\langle V \rangle$ and the transition condition $P_{\rm finite} <1$.  The average scales simply as the number of initial patches, so that if $\langle V \rangle$ is the average volume produced when starting with $V_\Lambda$, then $n \langle V \rangle$ is the volume produced when starting with $n V_\Lambda$ for $n > 1$.  Also, if $P_{\rm finite} =1$ for one initial patch of volume $V_\Lambda$, then $P_{\rm finite} = 1$ for $n$ initial patches since the patches evolve independently of one another.

%**************************************
%
%
%                    revised bacteria model

\subsection{Revised Bacteria Model} \label{revisedbacteriamodel}

Now we will modify the bacteria model approach taken in \cite{Dubovsky2009} to calculate the $\epsilon N_c$ corrections.  Start with a one dimensional lattice of sites with one bacterium on site $j$.  Let it replicate into $N_r$ copies, and let each copy hop independently with probability $p_j$ to site $j-1$ and probability $1-p_j$ to site $j+1$.  Now let each of these bacteria replicate and hop again, and continue the process.  In the model given in \cite{Dubovsky2009}, the hopping probability, $p_j$, and replication number, $N_r$, do not depend on the site $j$ and represent the motion of the field and expansion rate, respectively.  Here, we will allow $p_j$ and $N_r$ to depend on the bacteria's location on the lattice.  It can be shown that even if $p_j$ and $N_r$ depend on the site $j$, equation (\ref{apples}) remains valid, to leading order, with the only exception that $H$ and $\Omega$ become functions of $\phi$.  To see this, we will review the bacteria model proposed in \cite{Dubovsky2009} and its continuum limit, taking into account that $p_j$ and $N_r$ can depend on the site.  

To relate this model to the inflationary model, we demand that the bacteria model reproduces the stochastic equation for inflation given in \cite{Linde1986}:
\be \label{stochastic}
\pderiv{P}{t} = \pderivp{  \pderivp{ \frac{H^3}{8 \pi^2}P }{\phi} - \dot{\phi} P  }{\phi}.
\ee
This equation gives the probability $P( \phi , t )$ for the evolution of the field $\phi$ within a single Hubble patch.  This corresponds to the hopping of a single bacterium on the lattice.  Calling $P_{j,n}$ the discrete probability distribution for the bacterium to be at site $j$ at time step $n$, our bacteria model gives
\be
P_{j , n+1} = P_{j-1 , n} \left( 1 - p_{j-1} \right) + P_{j+1 , n}\, p_{j+1} \  .
\ee
Next let $P ( \phi_j , t_n ) =  \Delta \phi^{-1}_j P_{ j , n } $ (needed to preserve normalization through the limiting procedure) and $p ( \phi_j ) = p_{ j }$ which become functions of continuous variables as $\Delta \phi , \Delta t \rightarrow 0$.  We have let the discretization $\Delta \phi_j$ depend on site number in anticipation of  the matching condition.  Thus $\phi_j = \sum_{i=0}^j \Delta \phi_i$.  In the continuum limit $\Delta \phi (\phi_j) = \Delta \phi_j$ is also a function of a continuous variable.  After Taylor expanding for small $\Delta \phi$ and $\Delta t$ (for details see Appendix \ref{appendix2}), we get
\begin{align}
\Delta t \pderiv{P}{t}(\phi , t)  & = \frac{1}{2} \Delta \phi ( \phi)   \frac{\partial^2}{\partial \phi^2 } \Big( \Delta \phi  (\phi ) \, P(\phi , t)  \Big) \nonumber \\ & + \frac{\partial  \Delta \phi (\phi) }{\partial \phi}     \frac{\partial}{\partial \phi} \Big( \Delta \phi (\phi) \, P(\phi , t)\,  p(\phi)  \Big) + \frac{\partial}{\partial \phi} \Big( \Delta \phi(\phi) \, P(\phi)\,  ( 2 p(\phi) -1) \Big)  \ .
\end{align}
The continuum limit is reached by taking $\Delta \phi , \Delta t \rightarrow 0$ in such a way that $4 \pi^2 \Delta \phi (\phi)^2 = H(\phi)^3 \Delta t $, $ 2 p(\phi) -1  = - \dot{\phi}(\phi) \Delta t / \Delta \phi + (1/2) \partial \Delta \phi / \partial \phi$, and $\Delta t$ is independent of $\phi$.  We then follow the same steps as \cite{Dubovsky2009} to find the differential equation satisfied by $f$, keeping in mind the slightly different continuum limit.  The result is the same as equation (\ref{apples}) (up to corrections proportional to $\epsilon$) except that $H$ and $\dot{\phi}$ depend on $\phi$:
\be \label{fequation}
\frac{1}{2} \frac{\partial^2}{\partial \phi^2} f ( \phi; z) - \frac{2 \pi \sqrt{6 \Omega(\phi)} }{H(\phi)} \frac{\partial}{\partial \phi} f( \phi ; z) + \frac{12 \pi^2}{H(\phi)^2} f(\phi ; z) \log f(\phi ; z) =0 \ ,
\ee
subject to
\be
f(0; z) = e^{-z} \hspace{.3in} \text{and} \hspace{.3in} \frac{\partial}{\partial \phi} f(\phi ; z) \bigg|_{\phi_b} = 0 \ .
\ee
\begin{comment}
\be
\Delta t \pderiv{P}{t}(\phi , t)  = \frac{1}{2} \Delta \phi ( \phi)   \frac{\partial^2}{\partial \phi^2 } \Big( \Delta \phi (\phi ) P(\phi , t)  \Big) + \frac{\partial }{\partial \phi} \Delta \phi (\phi) \frac{\partial}{\partial \phi} \Big( \Delta \phi (\phi) P(\phi , t) p(\phi)  \Big) + \frac{\partial}{\partial \phi} \Big( \Delta \phi(\phi) P(\phi) ( 2 p(\phi) -1) \Big)  \ .
\ee
\end{comment}

%***********************************
%
%             order of approximation
%
%
%********************************

\subsection{Order of Approximation} \label{orderofapprox}
The bacteria model can describe a probability distribution $P(\phi)$ which comes from local Gaussian fluctuations whose variance and drift depend on position $\phi$.  However, slow roll inflation has non-Gaussianities of order the slow roll parameters, so these local fluctuations are not actually Gaussian.  Slow roll inflation also has metric fluctuations which are proportional to $\epsilon$ which are not captured by the bacteria model.  Thus, the bacteria model is an approximation which ignores terms of order $\epsilon$, so we should not expect to make any order $ \epsilon $ corrections by using it.  The bacteria model is also an approximation because it is a discrete model.  When passing to the continuum limit of the bacteria model we essentially take the large $N_c$ limit, so we should expect to miss contributions of order $1/N_c$.  

But, if we cannot keep terms of order $\epsilon$ or $1/N_c$, how will we make any corrections to the lowest order case?  The answer is that if inflation is long enough, i.e. there are many classical $e$-foldings $N_c$, the combination $\epsilon N_c$ can be of order one. In fact, if inflation is long enough, $\Omega$ may change substantially while $\epsilon$ is small.  In general, $\partial \Omega / \partial N \sim \epsilon \, \Omega$, so that $\Delta \Omega / \Omega \sim \epsilon N_c$ over a full inflationary trajectory.  Thus, we can expect to make corrections to the lowest order case at order $\epsilon N_c$, and we should ignore effects which are of order $\epsilon$ or $1/N_c$ only.  In this way, we do not obtain any information about the effect of local non-Gaussianities or metric fluctuations, but we do calculate slow roll effects which accumulate due to the length of inflation. In order for our calculation to be valid, the $\epsilon N_c$ corrections must dominate the other subleading corrections, i.e. $\epsilon N_c \gg \epsilon$ and $\epsilon N_c \gg 1/N_c$.  More specifically, our corrections to the volume scale like
\be \label{turntable}
\frac{ \delta \log \langle V \rangle  }{ \log \langle V \rangle } \sim \epsilon \, N_c \ .
\ee

Let us now verify that metric fluctuations give a contribution that is slow roll suppressed and not enhanced by $N_c$ with respect to the leading effect. The volume on the reheating surface can be expressed in $\zeta$ gauge as 
\be \label{volint}
V(\phi_r)=\int_{\phi=\phi_r} \sqrt{\hat{g}} =\int d^3 x \ a(t_r)^3 e^{3 \zeta(x,t_r)} \ ,
\ee
where $\hat g$ is the spatial metric induced on equal $\phi$ surfaces.  We can see that $\delta g_{00}$ and $\delta g_{0i}$ do not appear directly in the expression for the volume. They only affect the volume through their effect on the dynamics of $\zeta$. The leading effect comes from the cubic Lagrangian for $\zeta$, which is suppressed by $\epsilon$ with respect to the free Lagrangian \cite{Maldacena2002}. Thus the leading effect of the metric fluctuations on the dynamics of $\zeta$ comes through the one-loop correction to the power spectrum. This diagram scales like
\be
\delta \langle \zeta_k \zeta_k \rangle \sim \epsilon^2 P_{\zeta}(k) \ .
\ee
The factor of $\epsilon^2$ comes from the two insertions of the cubic vertex and the loop does not contribute a factor of $N_c$ because $\zeta$ correlators are time-independent at all loops \cite{Senatore:2012ya}. Going to real space contributes a factor of $N_c$ due to the time-dependence of the IR cutoff.  The amplitude $\Delta_\zeta$ defined by $P_\zeta (k ) = \Delta_\zeta / k^3$ is order one at the transition to eternal inflation, so the contribution to the $\zeta$ two-point function due to the metric fluctuations is
\be
\delta \langle \zeta(x) \zeta(x) \rangle \sim \epsilon^2 N_c \ .
\label{delta_zeta}
\ee
%The contribution to the volume from the change in $\zeta$ is
%\be
%\frac{\delta V}{V} \sim \delta \langle \zeta(x) \zeta(x) \rangle \sim \epsilon^2  N_c ,
%\ee
From~(\ref{volint}), we see that the leading effect on the volume due to $\delta \zeta$ is $ \delta  \log \langle V \rangle  / \log \langle V \rangle  \sim  \langle \delta \zeta^2 \rangle  / N_c $. Here we have used the fact that $\langle \zeta \delta \zeta \rangle =0$, since $\delta \zeta\sim \zeta^2$ in the cubic interaction. Thus the leading contribution of the metric fluctuations to the volume scales as
\be
\frac{ \delta \log \langle V \rangle }{ \log \langle V \rangle } \sim  \frac{ \langle \delta \zeta^2 \rangle }{N_c} \sim \epsilon^2 \ ,
\ee
which is subleading to the corrections considered in this paper which scale as in (\ref{turntable}).

At quadratic order, the interaction of the metric fluctuations with $\zeta$ are not generically suppressed by $\epsilon$. However, we can estimate the quadratic part of $\delta g_{00}$ with the Poisson equation. The contribution from $\delta g_{0i}$ is parametrically the same. Schematically, we have
\begin{equation}
H^2 M_{Pl}^2 \, \delta g_{00} \sim \delta \rho \sim \frac{1}{2} \left( \partial \delta \phi \right)^2 + V''(\phi) \delta \phi^2  \ .
\ee
Before estimating the effect of the metric fluctuations on the volume produced by inflation, we need to realize that if we take the expectation value on the right hand side, we get a non-zero value. This can be interpreted as a tadpole term of metric fluctuations that tells us that the tree-level background is not a solution of the equations of motion. Indeed, at this order we need to take into account the renormalization of the background by adding suitable counter-terms, similar to~\cite{Senatore:2009cf,Pimentel:2012tw}. If we do that then we can schematically write 
\begin{equation}
H^2 M_{Pl}^2 \, \delta g_{00} \sim \delta \rho - \langle \delta \rho \rangle   \ .
\ee
In calculating the variance of $\delta g_{00}$ from this, there are two types of terms to consider:
\begin{align}
&  \langle  \delta g_{00}^2 \rangle_{\delta \phi^2}=\frac{V''(\phi)^2}{H^4 M_{Pl}^4}  \Big\langle ( \delta \phi^2 - \langle \delta \phi^2 \rangle )^2 \Big\rangle  \ , \text{  and}  \label{metric_phi} \\
& \langle  \delta g_{00}^2 \rangle_{ \paren{\partial \delta \phi}^2}=\frac{1}{H^4 M_{Pl}^4} \Big\langle \left(  (\partial \delta \phi )^2 - \langle (\partial \delta \phi)^2 \rangle \right)^2 \Big\rangle  \label{metric_dphi}  \ .
\end{align}
Because $\delta \phi$ undergoes a random walk, $\langle ( \delta \phi^2 - \langle \delta \phi^2 \rangle )^2 \rangle \sim H^4 N_c$ is enhanced by the number of $e$-foldings.  However, $(\partial \delta \phi )^2$ does not undergo a random walk, since its two point function in momentum space is not scale invariant.  Thus, terms proportional to the variance of $(\partial \delta \phi )^2$ do not receive an enhancement by the number of $e$-foldings.  Putting this all together, equation~(\ref{metric_phi}) is proportional to $\epsilon^2 H^4 N_c/M_{Pl}^4$, and equation~(\ref{metric_dphi}) is proportional to $H^4/M_{Pl}^4$. 

We can find the effect of these quadratic metric fluctuations on the volume by estimating the back-reaction of  $\delta g_{00}$ on $\delta \phi$ through the Klein-Gordon equation. At linear order, the perturbed $\delta \phi$ scales like
\be
\delta (\delta \phi) \sim \delta g_{00} \delta \phi \ .
\ee
Because $\zeta$ and $\delta \phi$ are related by $\zeta=-H \delta \phi/\dot\phi$, 
\be
\frac{ \delta \log \langle V \rangle }{\log \langle V \rangle } \sim  \frac{ \langle  \delta \zeta^2 \rangle }{N_c}  \sim \frac{H^2}{\dot \phi^2 \, N_c} \langle  \delta \delta \phi^2 \rangle  \sim \frac{H^2}{\dot \phi^2 \, N_c} \langle \delta \phi^2 \, \delta g_{00}^2 \rangle \sim \frac{1}{ \Omega} \langle \delta g_{00}^2 \rangle
\ee
finally giving
%\be
%\frac{\langle \delta V \rangle }{\langle V \rangle} \sim \delta \langle \zeta^2 \rangle \sim \frac{H^2}{\dot \phi^2} \langle \delta g_{00}^2 \rangle  \langle \delta \phi^2 \rangle \sim \frac{ H^4 N_c}{\dot \phi^2} \langle \delta g_{00}^2\rangle \sim \frac{ N_c}{\Omega} \langle \delta g_{00}^2\rangle \ ,
%\ee
\be
\frac{\delta \log \langle  V  \rangle }{\log \langle V  \rangle } \Bigg|_{\delta \phi^2} \sim \frac{\epsilon^2 N_c}{\Omega} \frac{ H^4}{ M_{Pl}^4} \hspace{ .3in} \text{and} \hspace{.3in}\frac{\delta \log \langle V \rangle  }{\log \langle V \rangle  } \Bigg|_{(\partial \delta \phi)^2} \sim \frac{1}{\Omega } \frac{H^4}{M_{Pl}^4}  \ . 
\ee
Because $\Omega \approx 1$ at the transition to eternal inflation, these effects are suppressed with respect to the $\epsilon N_c$ effects considered in this paper. The metric fluctuations at cubic and higher order will be sourced by the lower order fluctuations, so the effect of the metric fluctuations on the volume can be neglected.

\section{Transition Condition for $f$} \label{transition}
Now that we have an equation that captures effects of order $\epsilon N_c$, we can study how the transition to eternal inflation changes. The results of~\cite{Dubovsky2009} show that to lowest order the transition to eternal inflation occurs at $\Omega=1$. In the revised model $\Omega$ varies with $\phi$, and we expect the transition to happen when $\phi_b$ is such that $\Omega ( \phi_b )-1=\mathcal{O} (1/N_c,\epsilon)$. This corresponds to a field range $\frac{\partial \Omega }{ \partial \phi  } \Delta \phi = 1/N_c$, which gives $ \Delta \phi \sim  H/ ( \sqrt{\Omega} \epsilon N_c) $. Assuming $\epsilon N_c \sim 1$, then near the transition point $\Delta \phi \sim H_*$, where starred quantities refer to the value at which $\Omega=1$. Thus we expect the transition to happen at 
\be \label{phirange}
\phi_b = \phi_* + \mathcal O(H_*) \ ,
\ee
so that the inflating region contains a Hubble-wide region in $\phi$-space that contains $\Omega\simeq 1$.

We can verify this suspicion by solving~(\ref{fequation}) for the generating function $f$. To find the transition point it suffices to study $f$ at $z=0$, so we will take $z=0$ for the rest of this section. To get some intuition for the solution for $f$, note that at leading order in slow roll, the differential equation~(\ref{apples}) can be thought of as describing the motion of a particle moving in the potential $U(f)=\frac{f^2}{4}\paren{\log{f^2}-1}$ depicted in Figure \ref{potential}. Here $\phi$ plays the role of time and there is an anti-friction term. In the revised model, the new differential equation~(\ref{fequation}) represents a particle moving in the same potential but now with a time-dependent pre-factor and anti-friction coefficient. The boundary conditions imply that the particle starts at $f=1$, the bottom of the potential well, at time $\phi_r$ and reaches some point $f_0$ at time $\phi_b$.
\begin{figure}[h!]
\begin{center}
\includegraphics[scale=.5]{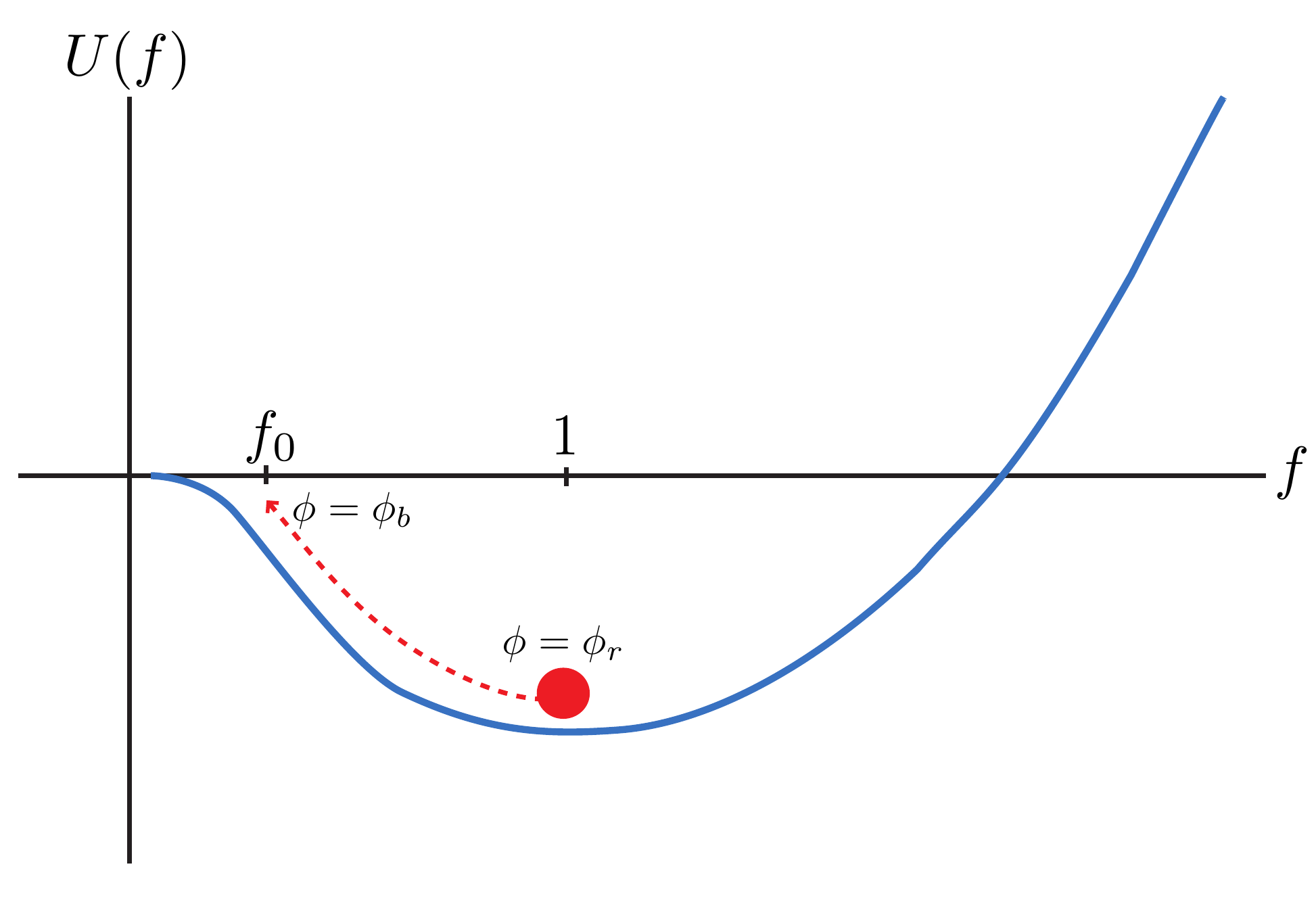}
\caption{The solution of (\ref{apples}) represents a particle moving in the potential $U(f)=\frac{f^2}{4} \left( \log{f^2 } -1\right) $. The particle starts at $f=1$ at time $\phi_r$ and is pushed up the potential by anti-friction to reach the point $f=f_0$ at time $\phi_b$. In the time-reversed process the particle starts at $f=f_0$ and rolls with friction to $f=1$.}\label{potential}
\end{center}
\end{figure}

%In \cite{Dubovsky2009}, the authors showed that in the constant $\Omega$ case it suffices to study $f$ at $z=0$ to see the transition to eternal inflation. This fixes the initial boundary condition to be $f(\phi_r)=1$. In the case $\Omega>1$, the particle is pushed to greater values of $f$ but the domain is restricted to $f\in[0,1]$ to normalize the probability, so it stays at $f=1$. For the $\Omega<1$ case the particle is pushed to smaller values of $f$ so it can climb up the potential until it reaches zero velocity. Thus for $\Omega>$, the solution is $f=1$ for all time and for $\Omega<1$, the starts at $f=1$ and is damped.  
%

In principle we could solve the revised equation (\ref{fequation}) for $f$. However, this is a boundary-value problem for a non-linear differential equation with time-dependent coefficients, which does not have an analytic solution and is numerically unstable. It will be simpler to consider the time-reversed process. In the time-reversed process, the boundary conditions require that the particle starts at $f_0$ with zero velocity and rolls with friction to $f=1$ at time $\phi_b-\phi_r$. Because of friction, this system has an attractor solution. Thus, even though the starting value of $f$ is not known, its value does not does not affect the behavior of the solution, and we are free to choose $f_0\approx 0$.  We cannot take it to be exactly zero because this is a stationary point that never reaches $f=1$.

How do we see the phase transition in the time-reversed solution for $f$? Let us first see the transition in the constant $\Omega$ case. The transition was studied in~\cite{Dubovsky2009} for the forward-time process, and the analysis is similar for the time-reversed process. The time-reversed equation for constant $\Omega$ is 
\be
\frac{1}{2}F''(\phi)+\frac{2\pi\sqrt{6 \Omega}}{H}F'(\phi)+\frac{2\pi^2}{H^2}F(\phi)\log{F(\phi)}=0 \ ,
\ee
where $F(\phi)=f(\phi_b-\phi)$. Around the potential minimum, $F \log (F)\approx F-1$ and the equation has an analytic solution, 
\be
F(\phi)=1+e^{-\frac{2\pi\phi}{H}\paren{\sqrt{\Omega}+\sqrt{\Omega-1}}}\paren{c_1+c_2 e^{\frac{4\pi\phi}{H}\sqrt{\Omega-1}}} \ ,
\ee
where $c_1$ and $c_2$ are constants that are fixed by the boundary conditions. It is clear that the solutions are oscillatory for $\Omega<1$ and non-oscillatory for $\Omega>1$. This marks the phase transition to eternal inflation. In the revised model $\Omega$ varies with $\phi$ so we will not be able to characterize the transition by a global value of $\Omega$. Instead, the behavior of $F$ will depend on how much of time is spent in the region where $\Omega<1$. Thus we need to solve for $F$ with different values of $\phi_b$ and select the critical $\phi_b$ that first makes $F$ oscillate. 

The time-reversed problem in the revised model is
\be
\frac{1}{2} \frac{\partial^2}{\partial \phi^2} F ( \phi) +\frac{2 \pi \sqrt{6 \Omega(\phi_b-\phi)} }{H(\phi_b-\phi)} \frac{\partial}{\partial \phi} F( \phi) + \frac{12 \pi^2}{H(\phi_b-\phi)^2} F(\phi ) \log F(\phi) =0 \ ,
\label{time-reversed}
\ee
with $F(0)\approx 0$ and $F'(0)\approx 0$. We will check the transition point for the sample potential $U=\frac{1}{2}m^2\phi^2$. Rewriting the system in terms of the dimensionless parameters $\varphi=\phi/\phi_*$ and $h_*=H_*/\phi_*$, (\ref{time-reversed}) becomes
\be
\frac{1}{2} \frac{\partial^2}{\partial \varphi^2} F ( \varphi) +\frac{2 \pi \sqrt{6} }{h_*(\varphi_b-\varphi)^3} \frac{\partial}{\partial \varphi} F( \varphi) + \frac{12 \pi^2}{h_*^2(\varphi_b-\varphi)^2} F(\varphi ) \log F(\varphi) =0 \ .
\label{nondim}
\ee
In this setup we must choose values for $\varphi_r$ and $h_*$.  The value of $\varphi_r$ does not affect the transition as long as $\varphi_b-\varphi_r$ is sufficiently large so that the solution reaches the attractor, so we will choose $\varphi_r$ to be the point at which $\epsilon=1$ to ensure that $\varphi_b-\varphi_r$ is as large as possible. With $\varphi_r$ fixed we can still vary $h_*$. For a given value of $h_*$, the behavior of $F$ will depend on the value of $\varphi_b$. Figure \ref{oscillations} shows the transition of $F$ from damped to oscillating. For each $h_*$ we find the critical $\varphi_b$ at which the maximum value of $F$ exceeds the asymptotic value $F=1$. With this critical $\varphi_b$ we calculate the field range inside the $\Omega=1$ region, $\Delta\varphi=\varphi_b-\varphi_*=\varphi_b-1$. Repeating this procedure for various values of $h_*$ gives a scatterplot of $\Delta\varphi$ versus $h_*$, which we fit to a line to extract the slope $\Delta\varphi/h_*=\Delta \phi /H_*$, the field range at transition normalized by $H_*$. Figure \ref{linfit} shows the data points along with the best fit line. The fit gives $\Delta \phi /H_*=1.166\pm0.006$. This confirms that the transition to eternal inflation occurs at $\Omega ( \phi_b )-1=\mathcal{O} (1/N_c,\epsilon)$. We control only order $\epsilon N_c$ deviations, so we find that the phase transition to eternal inflation still occurs at $\Omega=1$ for the revised model.

\begin{figure}[h!]
\begin{center}
\includegraphics[scale=.7]{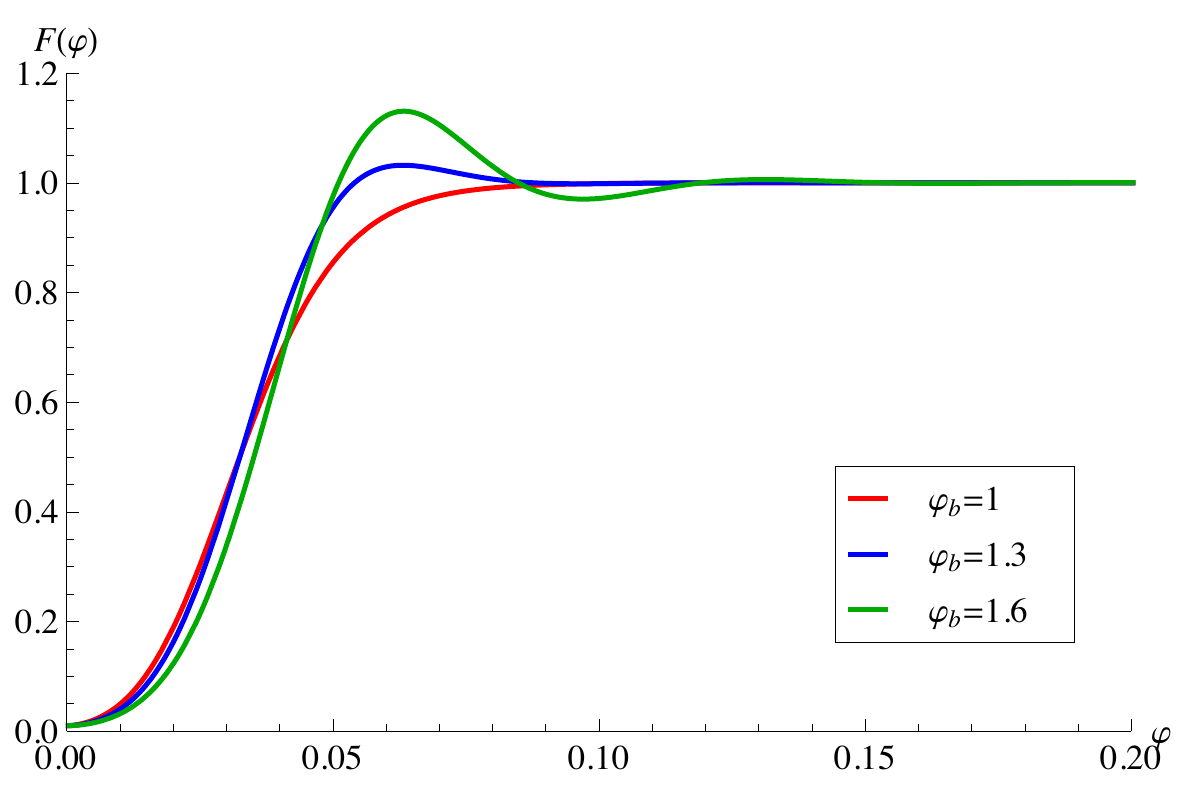}
\caption{Example plots of the solutions to (\ref{nondim}) for a particular value of the dimensionless Hubble parameter $h_*=0.1$ with varying values of $\varphi_b$. $\Omega<1$ for all $\varphi>1$ so we see that when the $\varphi_b$ is brought inside the $\Omega=1$ region the solution starts to oscillate.}\label{oscillations}
\end{center}
\end{figure}

\begin{figure}[h!]
\begin{center}
\includegraphics[scale=.7]{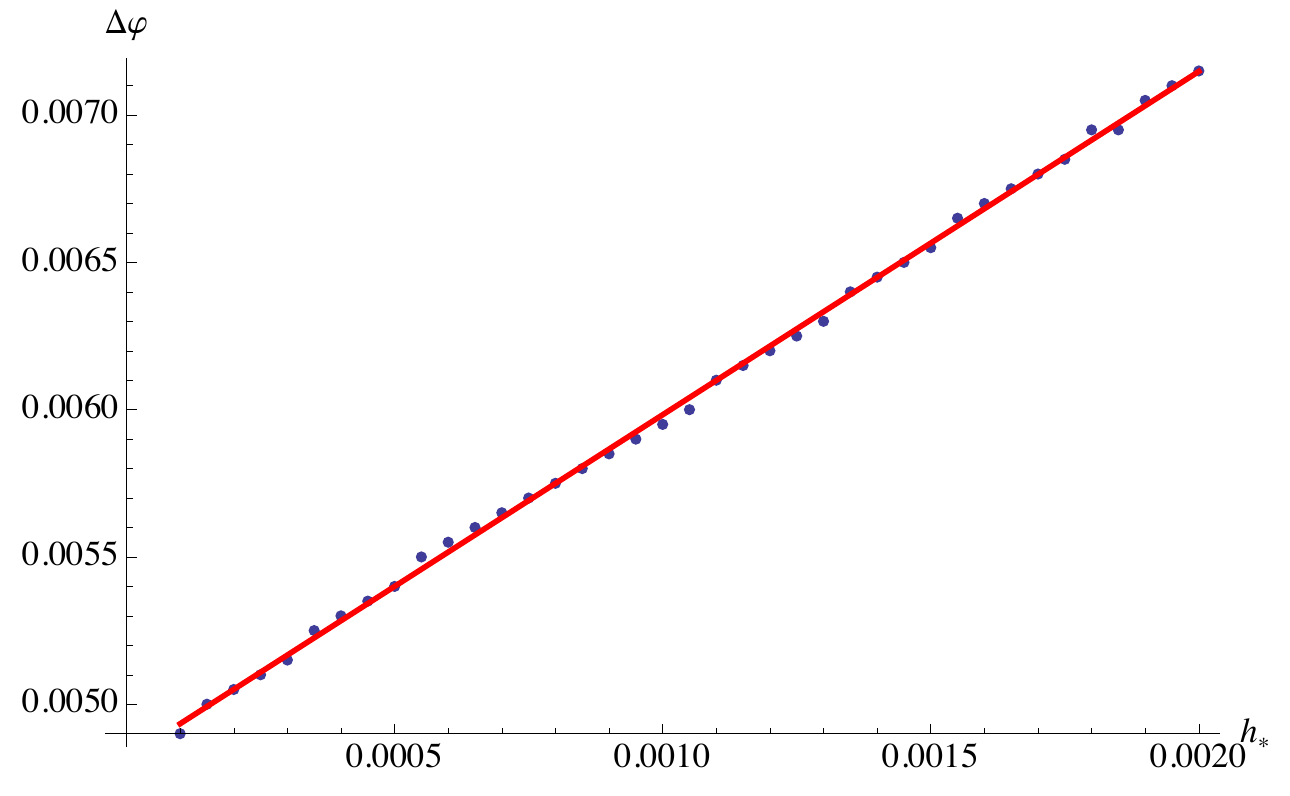}
\caption{A plot of the numerical values of $\Delta\varphi=\varphi_b-1$, where $\varphi_b$ is the point at which the numerical solution to (\ref{nondim}) ceases to oscillate. The $\Delta\varphi$ were evaluated for $h_*<0.002$, corresponding to $H_*/ M_{Pl} <0.01$. Also plotted is the linear fit to the data.}  \label{linfit}
\end{center}
\end{figure}

%********************************************
%%
%                       WKB

\section{WKB Approximation for Average Volume} \label{averagevolsec}

In addition to studying the transition condition by considering $f$ directly, we can also study the average volume.  Because equation (\ref{harrypotter}) for the average volume is linear, we can make progress in solving it analytically.  Below we use the WKB approximation to solve equation (\ref{harrypotter}).  As discussed in Appendix \ref{appendix}, it is sufficient to consider the solution in the classical regime $\Omega >1$, so we concentrate on that calculation below.  From now on, we will write $V \equiv \langle V \rangle$.

%************************************
%
%         classical regime
%
%

\subsection{Classical Regime: $\Omega >1$}

When deriving equation (\ref{harrypotter}), we have ignored terms of order $\epsilon$ and $1/N_c$ which come from non-Gaussianities in the inflaton field, corrections to the slow roll field equations, metric fluctuations, and the discrete nature of the bacteria model.  Thus, in the following approximations, we should also ignore these terms to stay self-consistent. 

To derive the WKB approximation, assume a solution of the form $-\dot{f}(\phi)=V(\phi) = e^{\sigma(\phi)} \tilde{V}(\phi)$ where $ \sigma ' = 2 \pi \sqrt{6 \Omega }/H$.  In terms of these variables, equation (\ref{harrypotter}) becomes
\be
 \label{energy} \tilde{V}''  =  w^2 \tilde{V}  \hspace{.1in} \text{ where } \hspace{.1in} 
w^2 = 24 \pi^2 \frac{ \Omega -1 }{H^2} \ ,
\ee 
$\sigma(\phi ) - \sigma( \phi_0) = \frac{ S_{\rm dS}(\phi_0) - S_{ \rm dS}(\phi) }{2}$, and we have ignored the $\sigma''$ term because it is proportional to the slow roll parameters. $S_{\rm dS} = 8 \pi^2 M_{Pl}^2 / H^2$ is the dS entropy in $3+1$ dimensions.  We can solve this by using WKB provided that $w' / w^2 \ll 1$, i.e. that $\partial( 1/ w ) / \partial \phi \ll 1$.  This approximation of course fails at the turning point $\Omega = 1$ where $w \rightarrow 0$.  The general solution away from $\Omega = 1$ (this precise range is discussed in Appendix \ref{appendix}) is then 
\begin{align}  \label{hertz}
 V(\phi)  &=e^{\sigma(\phi )} \left( A e^{ \int_{\phi_r}^{\phi} w(\phi') \, d \phi' } + B e^{ - \int_{\phi_r}^{\phi} w(\phi') \, d \phi'} \right) .
\end{align}
We have dropped the typical prefactor $1/\sqrt{w}$ because it is a subleading piece.  Next we need to impose the boundary conditions $V(\phi_r) = 1$ and $V'(\phi_b)=0$.  This gives a solution, valid as long as the solution doesn't cross the $\Omega = 1$ point,
\begin{equation} \label{wkb}
V(\phi) =  e^{ \frac{ S_{\rm dS}(\phi_r) - S_{\rm dS}(\phi)}{2}} \, \frac{ e^{I(\phi_r , \phi)} - \omega_+^2(\phi_b) e^{2 I( \phi_r , \phi_b)} e^{ -I ( \phi_r , \phi)}         }{    1 - \omega_+^2(\phi_b) e^{2 I(\phi_r, \phi_b)}   }  \hspace{.1in} \text{ for } \dot{\phi}<0 \ ,
\end{equation}
where $I( \phi_1 , \phi_2) = \int_{\phi_1}^{\phi_2} w(\phi') d \phi'$, and $\omega_{\pm} = \sqrt{ \Omega } \pm \sqrt{ \Omega - 1}$.  From this we can find the large barrier solution which reduces (\ref{wkb}) to 
\be \label{volume}
V_{\phi_b \rightarrow \infty} = e^{ \frac{S_{\rm dS}(\phi_r) - S_{\rm dS}(\phi) }{2}} \exp \left \{ - 2 \pi \sqrt{6}  \int_{\phi_r}^\phi \frac{\sqrt{ \Omega - 1}}{H}  \, d\phi' \right \} \ ,
\ee
or equivalently 
\be \label{volume2}
V_{\phi_b \rightarrow \infty} = \exp \left \{ 3 \int_0^{N_c} \frac{2}{1 + \sqrt{ 1 - 1/\Omega}} d N_c' \right \} \ .
\ee

%**************************
%
%            bound on the final volume
%            
%**********************

\subsection{A Bound on the Final Volume}

Let $\phi_*$ satisfy $\Omega ( \phi_* ) =1$, and let $N_c^*$ be the classical number of $e$-foldings that occurs when the field rolls from $\phi_*$ to $\phi_r$. Then the average volume produced just before eternal inflation is given by
\be  \label{volumebound}
V_{\rm{max}} = e^{ \frac{S_{\rm dS}(\phi_r) - S_{\rm dS}(\phi_*) }{2}} \exp \left \{ - 2 \pi \sqrt{6}  \int_{\phi_r}^{\phi_*} \frac{\sqrt{ \Omega - 1}}{H}  \, d\phi' \right \} \ ,
\ee
or equivalently 
\be \label{volumebound2} 
V_{\rm{max}} = \exp \left \{ 3 \int_0^{N_c^*} \frac{2}{1 + \sqrt{ 1 - 1/\Omega}} \, d N_c' \right \} \ .
\ee
Since we are only studying $\langle V \rangle$, we cannot tell the exact location of $\phi_b$ that determines the transition to eternal inflation.  But, as shown in Section \ref{transition}, the transition happens when $\phi_b$ is placed such that $\Omega( \phi_b) - 1 \sim \mathcal{O} (1/N_c, \epsilon)$.   This is expected since the results of \cite{Dubovsky2009} show that to lowest order, $\Omega = 1$ marks the transition, and that analysis did not capture effects of order $1/N_c$.  This corresponds to a field range $\phi_b - \phi_* \sim  \mathcal{O} (H_*)$ and says that we have an uncertainty in equation (\ref{volumebound}) which comes from replacing $\phi_*$ with $\phi_* \pm H_*$.  However, as shown in Appendix~\ref{appendix}, this does not affect the result at the order of approximation that we have considered.

There is another uncertainty in equation (\ref{volumebound}) which comes from the fact that we have only considered the WKB solution in the classical regime $\Omega > 1$.  However, if $\Omega ( \phi_b) < 1$, we should really consider the full WKB solution to solve across the turning point $\Omega = 1$.  This does not affect our calculation, since the exponential growth on the classical side $\Omega >1 $ dominates the contribution to the volume.  More details are provided in Appendix \ref{appendix}.

Equation (\ref{volumebound}) gives a large correction to the volume indeed!  The maximum possible value is $\log V_{\Omega = 1} = 6 N_c$ which occurs when $\Omega =1$ is constant.  The correction that the integral in (\ref{volume2}) makes is just the average value $(1/N_c) \int 1 / ( 1 + \sqrt{ 1 - 1/\Omega}) \, dN_c' \equiv r $ which is some number between $1/2$ and $1$.  In the case considered in \cite{Dubovsky2009}, the maximum volume that can be produced is $\log V = 6 N_c$.  By including corrections of order $\epsilon N_c$, we have found that the maximum volume is $\log V_{\rm max} = 6 \, r N_c$, an exponentially different volume.  This also shows that $V_{\rm{max}} \leq e^{S_{\rm dS}(\phi_r)/2}$, so that the universal bound is satisfied.  The only way to saturate the bound is to have $\Omega =1$ in the whole interval $(\phi_r,\phi_b)$, which is accomplished by a potential of the form $U(\phi) \sim \phi^{-2}$.  As shown in \cite{Dubovsky2009}, this is the maximum finite volume that can be produced, even during the eternal regime.

%%%%%%%%%%%%%%%%%%%%%%%
%
%
%            examples
%
%
%
%%%%%%%%%%%%%%%%%%%%%%

\subsection{Chaotic Inflation Examples}
%Then $\Omega = ( \phi_* / \phi )^{n+2}$, $\phi_* = \left( 2 \pi^2 n^2 M_{Pl}^6 / \mu^{4-n} \right)^{\frac{1}{n + 2}}$, $H^2 = \mu^{4-n} \phi^n / 3 M_{Pl}^2$, $N_c = \left( \phi_{\rm{start}}^2 - \phi_{\rm{end}}^2 \right) / 2 M_{Pl}^2 n$, and $\epsilon = - M_{Pl}^2 n^2 / 2 \phi^2$.  Then $\epsilon \ll 1 \Rightarrow M_{Pl}n  \ll \phi$ and $H^2 / M_{Pl}^2 \ll 1 \Rightarrow \mu^{4-n} \phi^n \ll M_{Pl}^4$.

In order to see explicitly how large these corrections are, we use the chaotic inflation scenario $U=~\mu^{4-n} \phi^n$ as an example \cite{Linde1983}.  In these cases $N_c = \left( \phi_{\rm{start}}^2 - \phi_{\rm{end}}^2 \right) / ( 2 M_{Pl}^2 n )$ and $\epsilon = - M_{Pl}^2 n^2 / (2 \phi^2)$.  Then the condition $\epsilon \ll 1$ implies $M_{Pl}n  \ll \phi$, and $H^2 / M_{Pl}^2 \ll 1$ implies $\mu^{4-n} \phi^n \ll M_{Pl}^4$.  For simplicity, we will take $n>0$ so that $\dot{\phi} < 0$ for $\phi > 0$.  The $n<0$ case is straightforward, but slightly different.  We will take the end of inflation to be when $\epsilon = 1$, $\phi_{\rm{end}}^2 = M_{Pl}^2 n^2 / 2$.  Then we can solve for $\phi_{\rm{start}}$ in terms of $N_c$.  Finally defining the dimensionless constant $\alpha = 2 M_{Pl}^2 n / \phi_*^2 \ll 1 $, we have from (\ref{volumebound})
\be
\log V_{\rm max} = 6 \int_0^{\frac{1}{\alpha} - \frac{n}{4} } \frac{d N_c }{ 1 + \sqrt{ 1 - \left( \alpha \left( N_c + n/4 \right) \right)^{1 + n/2} }} \ .
\ee
The integral can be done numerically.  For example, we obtain $\log V_{n = 2} = 6 \times 0.57 / \alpha$ and $ \log V_{n=4} = 6 \times 0.55 / \alpha$.  Comparing to the constant $\Omega$ case $\log V_{\Omega =1} = 6 N_c \approx 6 / \alpha$, we see that $\left( \log V_{\Omega = 1} - \log V_{2,4} \right) / \log V_{\Omega=1} \sim \mathcal{O}(1) = \mathcal{O} ( \epsilon N_c)$.  This remains true for larger $n$ as well.  Indeed, this is a substantial correction to the volume in the constant $\Omega$ case, changing it by a factor of about $e^{-3 N_c}$.

%%%%%%%%%%%%%%%%%%%%%%%%
%
%
%
%            conclusion
%
%
%%%%%%%%%%%%%%%%%%%%%%%%%%%

\section{Conclusion}
This paper has made three main accomplishments, all of which are valid when $\epsilon N_c$ corrections dominate $\epsilon$ and $1/N_c$ corrections.  Firstly, we have given the equation (\ref{fequation}) for the Laplace transform of the probability density $\rho(V )$ by studying a modified bacteria model of inflation.  We then performed two analyses of this equation.  The first, through numerical work, has determined that the condition for the transition to eternal inflation occurs when the barrier $\phi_b$ is placed such that $\Omega( \phi_b ) - 1 \sim \mathcal{O}(1/N_c)$.  Since our approach does not control corrections of order $1/N_c$, we conclude that the transition is consistent with $\Omega(\phi_b) = 1$.  The second analysis gave an analytic formula for the average volume $\langle V  \rangle (\phi) $ (\ref{volume}) as a function of the inflaton's initial value $\phi$.  This also allowed us to determine the maximum finite volume (\ref{volumebound}) that can be produced in any realization of the system, even in the eternal regime (remember that the eternal regime is defined by having a non-zero probability of infinite volume, so sometimes a finite volume is produced).  This analysis also established that the effects considered in this paper do not violate the universal entropy bound.  

By studying the revised bacteria model, where the hopping probability $p$ and the replication number $N_r$ depend on the site, we were able to calculate corrections to order $\epsilon N_c$.  The model we studied is a system with locally Gaussian fluctuations where the Gaussian shape depends on position $\phi$.  In this way, we do not capture local non-Gaussian features which start at order $\epsilon$, but we do capture global non-Gaussian features that are enhanced by $N_c$.  We have also ignored metric fluctuations which are of order $\epsilon$.  In order to capture these features with a bacteria model, significant changes would need to be made.  For instance, in order to determine the hopping probability in a non-Gaussian model at site $j$, a bacterium would have to remember its previous hops.  This problem is outside the scope of this paper.

%%%%%%%%%%%%%%%%%%%%%%

%       acknowledgements
%
%
%%%%%%%%%%%%%%%%%%%%%%%

\section*{Acknowledgements}
The authors would like to thank Leonardo Senatore and Giovanni Villadoro for extensive collaboration on this project.  ML acknowledges support from the NSF through the GRF program.  AP acknowledges the Gabilan Stanford Graduate Fellowship for support. Part of this work was done while visiting the CERN Theory Division, under support from NSF grant PHY-1068380.

%%%%%%%%%%%%%%%%%%%%%%%%%%%%%%%%
%%%%%%%%%%%%%%%%%%%%%%%%%%%%%%%

%                appendix

%%%%%%%%%%%%%%%%%%%%%%%%%%%

\appendix

\section{Continuum Limit Details} \label{appendix2}
The equation for the discrete probability distribution is 
\be
P_{j , n+1} = P_{j-1 , n} \left( 1 - p_{j-1} \right) + P_{j+1 , n}\, p_{j+1} \  ,
\ee
and the corresponding continuous quantities are defined by  $P ( \phi_j , t_n ) =  \Delta \phi^{-1}_j P_{ j  , n} $ and $p ( \phi_j ) = p_{ j}$.  Here, $\phi_j = \sum_{i=0}^j \Delta \phi_j$ and $t_n = n \, \Delta t$.  In the continuum limit, $\Delta \phi$ also becomes a function of $\phi$ by $\Delta \phi ( \phi_j) = \Delta \phi_{j }$.  This gives
\be \label{hipster}
\Delta \phi_j P( \phi_j , t_{n+1} ) = \Delta \phi_{j-1} P( \phi_{j-1} , t_n) \left( 1 - p(\phi_{j-1}) \right) + \Delta \phi_{j+1} P( \phi_{j+1} , t_n) p( \phi_{j+1}) \ .
\ee
Since we need an equation involving $\partial P / \partial t$ and $\partial^2 P / \partial \phi^2$, we must expand the left hand side to order $\Delta \phi \, \Delta t$ and the right hand side to $\Delta \phi^3$.  Thus we anticipate that $\Delta t \sim \Delta \phi^2$ in our matching condition.  We will expand functions $g(\phi)$ as
\be \label{g}
g( \phi_{j+1} ) = g( \phi_j + \Delta \phi_{j+1} ) = g(\phi_j ) + g'(\phi_j) \Delta \phi_{j+1} + \frac{1}{2} g''(\phi_j) \Delta \phi_{j+1}^2 + \mathcal{O}(\Delta \phi^3) \ ,
\ee
where $g' = \partial g / \partial \phi$. This includes expanding $\Delta  \phi_{j+1} = \Delta  \phi_j  + \Delta \phi_{j+1} \,  \Delta \phi '  (\phi_j)$.  From equation (\ref{g}) we can calculate 
\be
g( \phi_{j+1}) - g( \phi_{j-1} ) = g'( \phi_j ) \left( 2 \Delta \phi_j + \Delta \phi_j \frac{ \partial \Delta \phi}{\partial \phi} \Big|_{\phi_j} \right) + \mathcal{O}(\Delta \phi^3)
\ee
and 
\be
g( \phi_{j-1} ) = g(\phi_j) - g'( \phi_j ) \Delta \phi_j + \frac{1}{2} g''( \phi_j) \Delta \phi_j^2 + \mathcal{O}(\Delta \phi^3 ) \ .
\ee
Using equation (\ref{hipster}) this gives
\be
\Delta \phi \left( P + \Delta t \, \pderiv{P}{t} \right) = \Delta \phi \left(    P +  \frac{\partial}{\partial \phi} \Big( ( 2 p -1) \Delta \phi  P \Big)   +  \pderiv{\Delta \phi}{\phi} \frac{\partial}{\partial \phi} \Big( \Delta \phi  P p \Big) + \frac{\Delta \phi}{2} \frac{\partial^2}{\partial \phi^2} \Big( \Delta \phi  P \Big)      \right) \ .
\ee
Now we let 
\be \label{cont}
2 p - 1 = - \dot{\phi} \frac{\Delta t }{ \Delta \phi} + \frac{1}{2} \Delta \phi'  \hspace{.2in} \text{and} \hspace{.2in}  \frac{\Delta \phi(\phi)^2 }{\Delta t } = \frac{H(\phi)^3}{4 \pi^2} \ , 
\ee
and we reproduce (\ref{stochastic}).  

To obtain the equation for $f$, we proceed along the lines of \cite{Dubovsky2009}.  The bacteria model gives 
\be
f( \phi ; z) = \Big( \left( 1 - p(\phi) \right) f( \phi + \Delta \phi ; z ) + p(\phi) f( \phi - \Delta \phi ; z) \Big)^{N_r} \ 
\ee
and taking the continuum limit according to (\ref{cont}) with $N_r = 1 + 3 H(\phi) \Delta t$ gives equation (\ref{fequation}) up to a term which is proportional to $ \Delta \phi \, \Delta \phi ' f ' / \Delta t$.  This is proportional to $\epsilon$, so we do not include it in (\ref{fequation}).

%%%%%%%%%%%%%%%%%
%
%
%        more detailed analysis
%
%
%%%%%%%%%%%%%%%%%%%%

\section{Errors in Volume Expression} \label{appendix}
The formula given in (\ref{volumebound}) could have errors from three different sources, but we will show that they contribute subleading corrections.  The first error comes from our ignorance of the transition point.  We have checked that the transition occurs when $\Omega ( \phi_b ) -1 \sim \mathcal{O} ( 1/ N_c )$ which corresponds to a field range of $\phi_b - \phi_* \sim H_*$.  The second error comes from the fact that we have not used the full WKB solution which crosses the turning point $\Omega = 1$.  The third error comes from matching the two sides of the WKB solution across the turning point.

\subsection{Error in Ignorance of Transition Point}
Consider changing $\phi_* \rightarrow \phi_* \pm H_*$ in equation (\ref{volumebound}).  This produces a change 
\be
\delta \log V = - 2 \pi \sqrt{6} \frac{ \sqrt{\Omega} - \sqrt{\Omega - 1}}{H} \bigg|_{\phi_*} H_* =  \mathcal{O} (1) \ ,
\ee
 but $\log V \sim N_c$ in general, so this correction is $\mathcal{O} (1/N_c)$ which we don't control anyway.  Thus, our ignorance of the actual transition point is inconsequential to the order that we have calculated.  
%********************************
%
%         a closer analysis of the wkb solution
%
%%%%%%%%%%%%%%%%%%

\subsection{Error in Not Using Full WKB Solution} \label{d}

The regime relevant to the transition to eternal inflation is when $\phi_* \in (\phi_b , \phi_r)$ and the solution crosses from $\Omega >1$ to $\Omega < 1$.  This requires the use of Airy functions which sew the two WKB solutions together.  For a potential which gives $\dot{\phi} < 0$, let $\phi_*$ be defined by $\Omega(\phi_*)=1$ as usual.  Then $\Omega < 1$ for $\phi > \phi_*$.  We will take $S(\phi)\equiv S_{\rm dS}(\phi)$ for brevity.  Then the solution which satisfies the boundary conditions is 
\be \label{wkbsol}
V(\phi) = \begin{cases} V_L(\phi) \equiv e^{ \frac{ S(\phi_r) - S(\phi)}{2}} F(\phi_*,\phi_r,\phi_b) \left[ t(\phi_*,\phi_b) e^{I( \phi_* , \phi )} + 2 e^{-I(\phi_* , \phi)} \right]  \text{  for  } \phi< \phi_* - \Delta \phi \\
V_R(\phi) \equiv e^{ \frac{ S(\phi_r) - S(\phi)}{2}} F(\phi_*,\phi_r,\phi_b) \\
 \hspace{1in} \times \left[ \left( 1 - i t(\phi_*,\phi_b) \right) e^{ i J(\phi_* , \phi) + i \frac{\pi}{4}} + \left( 1 + i t(\phi_*,\phi_b) \right) e^{ -i J(\phi_* , \phi) - i \frac{\pi}{4}} \right] \\
 \hspace{4in} \text{  for  } \phi > \phi_* + \Delta \phi
\end{cases}
\ee
where 
\be
F(\phi_*,\phi_r, \phi_b) =\frac{ e^{I(\phi_* , \phi_r)}    }{  t(\phi_* , \phi_b) e^{ 2I(\phi_*,\phi_r)}+2  } \ ,
\ee
\be
t(\phi_* , \phi_b) = i \frac{ 1 + i \tilde{ \omega}_+(\phi_b)^2 e^{2 i J(\phi_*,\phi_b)}}{1 - i \tilde{ \omega}_+(\phi_b)^2 e^{2 i J(\phi_*,\phi_b)}} \ , 
\ee
\be
J(\phi_1 , \phi_2) = 2 \pi \sqrt{6} \int_{\phi_1}^{\phi_2} \frac{ \sqrt{ 1- \Omega } }{ H } \, d \phi'
\ee
and $ \tilde{\omega}_+ ( \phi) = \sqrt{\Omega (\phi) } +  i \sqrt{ 1 - \Omega (\phi) }$.  $I(\phi_1,\phi_2)$ is defined below equation (\ref{wkb}).  This is the correct solution away from the turning point $\Omega =1$.  We will show in Appendix \ref{airy} how close to $\phi_*$ this equation can be used (i.e. how large $\Delta \phi$ is), and that the matching provided by the Airy functions is accurate to the order that we need.

The volume at the barrier position is given by 
\be \label{vb}
V(\phi_b)  =\delta \cdot  e^{( S_r - S_b) /2} \frac{ \tan \left( J_b + \alpha_b + \pi /4 \right)  }{ \tan \left( J_b + \alpha_b + \pi /4 \right) - \delta^2 / 2 } \left[ \cos \left( J_b + \pi / 4 \right) - \frac{ \sin \left( J_b + \pi / 4 \right) }{ \tan \left( J_b + \alpha_b + \pi / 4 \right)} \right] \ ,
\ee
where $S_r \equiv S(\phi_r)$, $S_b \equiv S(\phi_b)$, $J_b \equiv J(\phi_*, \phi_b)$, $\alpha_b \equiv \alpha(\phi_b)$, 
\be
\delta = \exp \left( 2 \pi \sqrt{6} \int_{\phi_*}^{\phi_r} \frac{ \sqrt{ \Omega - 1} }{ H } \, d \phi \right) \ll 1 \ ,
\ee
and
\be
\alpha(\phi) = \arctan \left( \sqrt{ \frac{1}{\Omega(\phi)} - 1 } \right) \ .
\ee

The first thing to notice is that (\ref{vb}) is $\delta \cdot  e^{( S_r - S_b) /2}$ (the solution that we gave in equation (\ref{volumebound})) times factors that are usually order one.  The only time that corrections are important is when $  \tan \left( J_b + \alpha_b + \pi /4 \right) \rightarrow \delta^2 / 2 $ and the solution starts to diverge.   However, we will show that this happens over a field range $\Delta \phi/H \sim e^{-N_c}$ and so is certainly not under control in our calculation.

The volume is positive until it diverges at a value of $\phi_b \equiv \phi_{div}$ such that $ \tan \left( J_b + \alpha_b + \pi /4 \right) = \delta^2/2$, or, since $\delta \ll 1$, at 
\be
  J (\phi_*, \phi_{div} ) + \alpha(\phi_{div}) = \frac{\delta^2}{2} + \frac{3 \pi}{4} \ .
\ee
Expanding $J(\phi_*, \phi)$ near $\phi \approx \phi_{div}$ gives 
\begin{align}
J(\phi_*,  \phi - \phi_{div} ) & \approx J ( \phi_*, \phi_{div} ) +  \partial_\phi J( \phi_* , \phi)\big|_{\phi_{div}} (  \phi - \phi_{div}  ) + \dots  \\
& \equiv  J (\phi_* , \phi_{div} ) + A \cdot \Delta \phi 
\end{align}
where $A =  \partial_\phi J( \phi_* , \phi)\big|_{\phi_{div}}$ and $\Delta \phi = \phi - \phi_{div}$.  Under this approximation, for $\phi_b \approx \phi_{div}$, the volume becomes 
\be
V(\phi_b) \approx e^{\Delta S  / 2} \frac{\delta^3 }{2} \frac{1}{A \Delta \phi} \left( - \cos \left( - \alpha_b + A \Delta \phi + \delta^2 / 2 \right) + \frac{2}{\delta^2} \sin \left( - \alpha_b + A \Delta \phi + \delta^2 / 2 \right) \right) \ .
\ee
When the arguments of the trigonometric functions are small
\begin{align}
V(\phi_b) & \approx  \frac{ \delta^3 }{2} \frac{e^{\Delta S  / 2}}{A \Delta \phi}   \left(  - 1   + \frac{2}{\delta^2} \left(  A \Delta \phi + \frac{\delta^2 }{ 2} -  \alpha_b    \right) + \dots \right) \\
& = \frac{ \delta^3 }{2} \frac{e^{\Delta S  / 2}}{A \Delta \phi}  \left( - \frac{2 \alpha_b }{\delta^2}  + \frac{ 2 A \Delta \phi }{\delta^2} + \dots  \right) \ ,
\end{align}
where $\dots$ represent terms of order $\delta^2 - \alpha_b$ and $(\delta^2 - \alpha_b )^k / \delta^2$ for $k>1$.  This simplifies to 
\be
V( \phi_b ) \approx e^{ ( S_r - S_b ) /2 } \delta \left[ \frac{1}{2 \pi \sqrt{6} \sqrt{ \Omega(\phi_{div})}} \, \frac{H(\phi_{div})}{\phi_{div} - \phi_b} + 1 \right] \ .
\ee

Since generally $V \sim e^{N_c}$, this does not make a correction to our order of approximation until
\be \label{divwidth}
\frac{ \phi_{div} - \phi_b}{H(\phi_{div}) } = \frac{e^{-N_c}}{2 \pi \sqrt{ 6 \Omega(\phi_{div})} } \sim e^{-N_c} \ ,
\ee
since near the transition, $\Omega \approx 1$.  We certainly do not have control over this kind of non-perturbatively small number, so the divergence is probably just an artifact of our approximation (like, for instance, imposing the exact boundary conditions below equation (\ref{hertz})).  For the final volume, we should take the one which is approximately constant over field ranges of order $\Delta \phi_{\rm{tot}} / N_c$, which is just 
\be
V_{\rm{max}} = e^{(S_r - S_b)/2} \delta \ . 
\ee
Since $\delta < 1$, this volume also satisfies the universal bound.  This is an upper bound on the average volume, as discussed after equation (\ref{oldvolume}).

%%%%%%%%%%%%%%%%%%%%%%%%%%%%%%%
%
%
%             matching
%
%
%%%%%%%%%%%%%%%%%%%%%%%%%%%%%%%%%%%%
\subsection{ Error in Matching} \label{airy}
The last errors could come from matching the WKB solutions across the turning point $\phi_*$.  We have neglected the factor of $1/ \sqrt{w(\phi)} $ in (\ref{hertz}), but for $\phi \approx \phi_*$ this factor becomes large and would contribute to our solution.  This produces a different volume $\tilde V_{(2)} = \tilde V_{(1)} A_{\rm WKB}$, where $\tilde V_{(1)}$ is the first order WKB solution in (\ref{energy}).  To estimate how the error $A_{\rm WKB}$ depends on the distance $\Delta \phi$ away from $\phi_*$, consider the next order WKB solution

\be
\tilde{V}_{(2)} ( \phi ) = \exp \left\{ \int_{\phi_r}^{\phi} d \phi' \left( - w(\phi') - \frac{1}{2} \frac{\partial_{\phi'} w(\phi')}{w(\phi')} \right) \right\} \ .
\ee
This is a good approximation until $w \approx 0$ over a long enough region such that the correction becomes large.  Let $\phi_s$ satisfy $\partial_\phi w( \phi_s) / w(\phi_s) \approx w(\phi_s)$.  Then, even $\tilde V_{(2)} ( \phi_s ) $ is an accurate approximation because $\partial_\phi w / w^2 \ll 1 $ in most of the integration region.  This means that $A_{\rm WKB} \sim 1$ for any $\phi < \phi_s$.  For $ \phi_s < \phi < \phi_*$, we can estimate the error as follows.  We can write
\be
\tilde V_{(2)} ( \phi ) = \exp \left\{ - \int_{\phi_r}^\phi d \phi' w(\phi') - \frac{1}{2} \int_{\phi_r}^{\phi_s} d \phi' \frac{ \partial_{\phi'} w(\phi') }{w(\phi')}   - \frac{1}{2} \int_{\phi_s}^{\phi} d \phi' \frac{ \partial_{\phi'} w(\phi') }{w(\phi')}     \right\} \ . 
\ee
The first two integrals in the above expression give something of order $N_c$, and the correction of the second integral to the first is small.  The approximation only fails when $\phi \rightarrow \phi_*$ and the third integral becomes large.  Thus, the error for taking $\phi = \phi_* - \Delta \phi$ is given by 
\be
A_{\rm WKB} = \exp \left\{      - \frac{1}{2} \int_{\phi_s}^{\phi_* - \Delta \phi} d \phi \, \frac{ \partial_{\phi} w(\phi) }{w(\phi)}     \right\} = \sqrt{ \frac{ w(\phi_s)}{w(\phi_* - \Delta \phi)}}  \hspace{.2in} \text{ for} \hspace{.2in} \Delta \phi< \phi_* - \phi_s \ . 
\label{awkb}
\ee
To estimate the error we just need to find $\Omega(\phi_s)$, which is when $1 \sim H  ( \Omega - 1)^{-3/2} \partial_\phi \Omega  - \partial_\phi H$.  Using $\partial_\phi \Omega \sim \epsilon \sqrt{\Omega}/ H$, this means that $\Omega(\phi_s) = 1 + \mathcal{O}( \epsilon^{2/3})$ and $(\phi_* - \phi_s)/H_* \sim \epsilon^{-1/3}$.  Using this $\phi_s$, (\ref{awkb}) becomes
\begin{align} \label{purple}
 \sqrt{ \frac{ w(\phi_s)}{w(\phi_* - \Delta \phi)}}  =\paren{ \frac{H(\phi_* - \Delta \phi)}{ H(\phi_s) } }^{1/2} \paren{  \frac{ \Omega(\phi_s) - 1}{ \Omega( \phi_* - \Delta \phi) - 1}         }^{1/4} & \sim \paren{ \frac{ \epsilon^{2/3}}{  | \partial_\phi \Omega( \phi_*) |  \Delta \phi}}^{1/4} \nonumber  \\
 & \sim \left( \frac{H_*}{ \epsilon^{1/3} \Delta \phi} \right)^{1/4} \ .
\end{align}
Thus, the total expression for the WKB error is
\be
A_{\rm WKB} \sim \begin{cases} \hspace{.25in} 1 & \text{for} \hspace{.1in} \Delta \phi > \phi_* - \phi_s \\  \left( \frac{H_*}{ \epsilon^{1/3} \Delta \phi} \right)^{1/4} & \text{for} \hspace{.1in} \Delta \phi < \phi_* - \phi_s   \end{cases} \ . 
\ee 
If $A_{\rm WKB}$ becomes of order $e^{ \epsilon N_c^2}$, then we need to include it.  However, we match to the Airy functions at some $\Delta \phi$ and then use those solutions, so we can neglect the factor of $1 / \sqrt{w(\phi)}$ as long as $\Delta \phi$ is large enough.  

Of course, $\Delta \phi$ cannot be too large because that would compromise the validity of the Airy solutions.  Near $\phi_*$, equation (\ref{energy}) becomes 
\be \label{sun}
\frac{ \partial^2 \tilde V^{  A}_{(2)} }{ \partial \phi^2} - \frac{ \partial w^2}{\partial \phi} \Big|_{\phi_*} \Delta \phi  \, \tilde V^{ A}_{(2)} - \frac{1}{2} \frac{ \partial^2 w^2 }{\partial \phi^2} \Big|_{\phi_*} (\Delta \phi)^2 \tilde V^{ A}_{(2)} = 0
\ee
to second order in $\Delta \phi$.  Typically, we drop the term proportional to $(\Delta \phi)^2$ and then the solution is a linear combination of the Airy functions,  $ \tilde V^{ A}_{(1)} = a \,  Ai( z) + b \,  Bi(z)$ where $z = ( \partial w^2 / \partial \phi |_{\phi_*} )^{1/3} \Delta \phi \sim \epsilon_*^{1/3} \Delta \phi / H_*$.  Including the term proportional to $(\Delta \phi)^2$ gives a different solution $\tilde V^{ A}_{(2)} = \tilde V^{A}_{(1)} A_{\rm Airy}$.  To estimate the error, let $A_{\rm Airy} = 1 + \gamma $ for $\gamma \ll 1$.  Then equation (\ref{sun}) becomes
\be
\frac{ \partial^2 \paren{ \gamma \, \tilde V^{  A}_{(1)}} }{ \partial \phi^2} - \frac{ \partial w^2}{\partial \phi} \Big|_{\phi_*} \Delta \phi  \, \gamma \,   \tilde V^{ A}_{(1)} - \frac{1}{2} \frac{ \partial^2 w^2 }{\partial \phi^2} \Big|_{\phi_*} (\Delta \phi)^2 \tilde V^{ A}_{(1)} = 0 \ . 
\ee
This correction becomes important when the third term becomes of order the second term, which gives $\gamma \sim \Delta \phi \, \partial^2_\phi \,  w^2 / \partial_\phi \, w^2 |_{\phi_*}$.  Since $A_{\rm Airy} \approx e^\gamma$, the next order correction to the volume due to the Airy function is
\be
A_{\rm Airy} \sim \exp \left\{ \frac{ \epsilon_*  \Delta \phi}{  H_* } \right\} \ . 
\ee
Now, all we have to do is choose a $\Delta \phi$ such that $z \gg 1$ and $A_{\rm WKB}$ and $A_{\rm Airy}$ are both much smaller than $e^{ \epsilon N_c^2}$.  For example, we can make $A_{\rm WKB}$ and $A_{\rm Airy}$ smaller than $e^1$ by choosing $\Delta \phi$ in the range
\be
e^{-4  } N_c^{1/3} \lesssim \Delta \phi / H_* \lesssim  N_c \ ,
\label{delta_phi}
\ee
where we have taken $\epsilon_* \sim N_c^{-1}$. We also need $z \gg 1$ to be able to match the Airy functions in the asymptotic limit, which implies $\Delta \phi / H_*\gg N_c^{1/3}$. Since $N_c\gg1$ we can choose $\Delta \phi /H_* \sim N_c^{2/3}$ and satisfy all the constraints.  This gives $A_{\rm WKB} \sim 1$ and $A_{\rm Airy} \sim e^{N_c^{-1/3}}$.  We already neglect terms of this order in the derivation of the equation for the average volume, so these effects are clearly negligible.

\bibliographystyle{unsrt}      %doesn't sort bib entries, so they show up in the order that they appear
\bibliography{slow_roll_volume_final.bib}

\flushbottom                                                           %LMS

\end{document}